\documentclass[graybox]{svmult}

\usepackage{mathptmx}       % selects Times Roman as basic font
\usepackage{helvet}         % selects Helvetica as sans-serif font
\usepackage{courier}        % selects Courier as typewriter font
\usepackage{makeidx}         % allows index generation
\usepackage{graphicx}        % standard LaTeX graphics tool
\usepackage{multicol}        % used for the two-column index
\usepackage[bottom]{footmisc}% places footnotes at page bottom

\def\aap{Astron. Astrophys.}
\def\apj{Astrophys. J.}
\def\apjl{Astrophys. J. Lett.}
\def\solphys{Solar Phys.}
\def\ssr{Space~Sci.~Rev.}
\def\pasj{PASJ}

\newcommand{\ck}{c_{\mathrm k}}

\newcommand{\vap}{v_{\mathrm{Ap}}}
\newcommand{\rhop}{\rho_{\mathrm{p}}}
\newcommand{\rhoc}{\rho_{\mathrm{c}}}
\newcommand{\cp}{c_{\mathrm{p}}}

\makeindex             % used for the subject index
\begin{document}

\title*{Inversion of physical parameters in solar atmospheric seismology}
% Use \titlerunning{Solar atmospheric seismology} for an abbreviated version of
\titlerunning{Inversion of physical parameters in solar atmospheric seismology}
\author{I\~nigo Arregui}
\institute{I\~nigo Arregui \at Departament de F\'{\i}sica, Universitat de les Illes Balears, E-07122 Palma de Mallorca, Spain \\\email{inigo.arregui@uib.es}}
\maketitle

\abstract*{Magnetohydrodynamic (MHD) wave activity is ubiquitous in the solar atmosphere. 
MHD seismology aims to determine difficult to measure physical parameters in solar atmospheric magnetic and plasma structures by a combination of observed and theoretical properties of MHD waves and oscillations. This technique, similar to seismology or helio-seismology, demands the solution of two problems. The direct problem involves the computation of wave properties of given theoretical models. The inverse problem implies  the calculation of unknown physical parameters, by means of a comparison of observed and theoretical wave properties. Solar atmospheric seismology has been successfully applied to different structures such as coronal loops, prominence plasmas, spicules, or jets.  However, it is still in its infancy. Far more is there to come. We present an overview of recent results, with particular emphasis in the inversion procedure.}

\abstract{Magnetohydrodynamic (MHD) wave activity is ubiquitous in the solar atmosphere. 
MHD seismology aims to determine difficult to measure physical parameters in solar atmospheric magnetic and plasma structures by a combination of observed and theoretical properties of MHD waves and oscillations. This technique, similar to seismology or helio-seismology, demands the solution of two problems. The direct problem involves the computation of wave properties of given theoretical models. The inverse problem implies  the calculation of unknown physical parameters, by means of a comparison of observed and theoretical wave properties. Solar atmospheric seismology has been successfully applied to different structures such as coronal loops, prominence fine structures, spicules, or jets. However, it is still in its infancy. Far more is there to come. We present an overview of recent results, with particular emphasis in the inversion procedure.}

%%%%%%%%%%%%%%%%
%%%%%%%%%%%%%%%%
\section{Introduction}
%%%%%%%%%%%%%%%%
%%%%%%%%%%%%%%%%

The term solar atmospheric seismology refers to the study of the physical conditions in solar atmospheric magnetic and plasma structures by the study of the properties of waves in those structures. The aim is to increase our knowledge about the complicated structure and dynamics of the solar atmosphere. The reason why  solar atmospheric structures can be probed from their oscillations is that the properties of these oscillations are entirely determined by the plasma and magnetic field properties. This  remote diagnostics technique  was first suggested by \cite{uchida70} and \cite{roberts84}, in the coronal context, and by \cite{roberts94} and \cite{tandberg-hanssen95} in the prominence context.  Solar atmospheric seismology  has experienced a great advancement in the last decade, made possible by the increase in the quantity and quality of wave activity observations obtained from space-borne observatories (TRACE, Hinode, SDO), and the refinement of theoretical MHD wave models. 

%We are now at a moment in which we can aim at comparing theory and observations in a confident way and  obtain reliable information about the physical conditions in the solar atmosphere and the nature of the different manifestations of solar atmospheric wave dynamics.

\begin{figure}[t]
\centering
  \includegraphics[height=6cm,width=8.5cm]{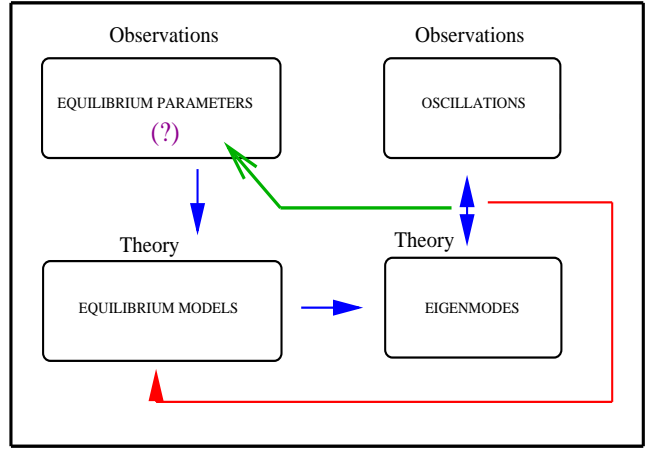} 
   \caption{The method of MHD seismology.}
   \label{fig:sistematic}
\end{figure}

The method of MHD seismology is displayed in Fig.~\ref{fig:sistematic}. Observations of solar coronal magnetic and plasma structures provide us with information that can be used to construct theoretical models. Observations also provide us with measurements of certain wave properties, such as periods, damping times, or additional parameters, such as mass flow speeds. By analyzing the wave properties of given theoretical models (direct problem) these can be compared to the observed wave properties, and difficult to measure physical quantities can be obtained (inverse problem).

This paper presents an overview of recent results obtained from the application of MHD seismology techniques to infer physical properties in solar atmospheric structures.  Some representative examples are selected. Our emphasis will be on the different inversion techniques that have been used.

%%%%%%%%%%%%%%%%%%%%%%%%%%%%%%%%%
%%%%%%%%%%%%%%%%%%%%%%%%%%%%%%%%%

%End this paragraph with references to reviews on observations \cite{aschwanden03a,nakariakov05,banerjee07}.
 
%%%%%%%%%%%%%%%%
%%%%%%%%%%%%%%%%
\section{Seismology Using the Period of Oscillations}\label{period}
%%%%%%%%%%%%%%%%
%%%%%%%%%%%%%%%%

Quickly damped transverse coronal loop oscillations where first reported 
by \cite{aschwanden99} and \cite{nakariakov99}. They are interpreted as the fundamental MHD kink mode of a magnetic flux tube. Using this interpretation \cite{nakariakov01} performed
the first modern seismology application, determining the magnetic field strength in a coronal loop.
Their analysis was based on the observational determination of the period ($P$) of the oscillation and length ($L$) of the loop, that lead to an estimate of the phase speed 

\begin{equation}
\frac{\omega}{k}=\frac{2L}{P},
\end{equation}
with $\omega$ the frequency and $k$ the longitudinal wavenumber. Theory for standing kink waves in the long wavelength approximation  tells us that the phase speed can be approximated by the so-called kink speed, $\ck$, which in terms of the physically relevant quantities  reads

\begin{equation}\label{kink}
\frac{\omega}{k}\approx \ck \equiv v_{\rm Ai}\left [\frac{2\zeta}{1+\zeta}\right]^{1/2}.
\end{equation}
In this expression, $v_{\rm Ai}$ is the Alfv\'en speed in the loop and $\zeta=\rho_{\rm i}/\rho_{\rm e}$ the ratio of internal to external densities. Notice that both quantities are unknown, hence no unique solution can be obtained from the observed period alone. Further progress with algebraic relation (\ref{kink}) requires the consideration of the density contrast a a parameter, to obtain the Alfv\'en speed. Then the magnetic field strength is determined as $B=(4\pi \rho_i)^{1/2}v_{\rm Ai}$. By considering loop number densities in the range
[1-6]$\times 10^{9}$ cm$^{-3}$, magnetic field strengths in between 4 and 30 G are obtained. The method outlined by \cite{nakariakov01} for single mode seismology of coronal loops has been subsequently employed using better observations and more accurate data analysis techniques. Some relevant analyses can be found in \cite{vandoorsselaere08,verwichte09,white12}.

Prominence fine structures also display transverse oscillations.  A recent study  by \cite{lin09} follows the same inversion procedure as \cite{nakariakov01} applying it  to propagating transverse thread oscillations.
A fundamental difference with respect to the coronal loop case is that, in the limit of high density contrast typical of prominence plasmas, the ratio $\rho_{\rm i}/\rho_{\rm e}$ is very large and the ratio $\ck^2 /v_{\rm Ai}^2$ is almost independent from it. The kink speed can then be approximated by 
\begin{equation}
\ck \approx \sqrt{2} v_{\rm Ai}. \label{kinks2}
\end{equation}
Reference \cite{lin09} assumed that thread oscillations observed from the H$\alpha$ sequences were the result of a propagating kink mode, which implies that the measured phase velocity, $\cp$, is equal to the kink speed. Then, the prominence thread  Alfv\'en speed ($\vap$) can be computed from
\begin{equation}
\vap \approx \frac{\cp}{\sqrt{2}}.
\end{equation}
The inferred values of $\vap$ for ten selected threads show that the physical conditions in different threads were very different in spite of belonging to the same filament.  Once the Alfv\'en speed in each thread was determined, the magnetic field strength could be computed when a value for the thread density was adopted. For the analyzed events, and considering a typical value for the prominence density, $\rho_{\rm i}=5\times10^{-11}$~kg~m$^{-3}$, magnetic field strengths in the range 0.9\,--\,3.5~G were obtained.

The widespread use of the concept of period ratios as a seismological tool (reviewed by \cite{andries09}) has been remarkable in the context of coronal loop oscillations. The idea was first put forward by \cite{andries05b,goossens06} as a means to infer the coronal density scale height using multiple mode oscillations in coronal loops embedded in a vertically stratified atmosphere. In coronal loop seismology, the ratio of the fundamental mode period to twice that of its first overtone in the longitudinal direction ($P_1/2P_2$) mainly depends on the density structuring along magnetic field lines. It is equal to unity in longitudinally uniform tubes, but is smaller than  one when density stratification is present (magnetic field stratification has the opposite effect as we discuss in Sect.~\ref{spatial}).  It can therefore be used as a diagnostic tool for the coronal density scale height, since there is a one-to-one relation between density stratification and period ratio. Using observational estimates for two simultaneous multiple mode observations by \cite{verwichte04}, \cite{andries05b} find that both observations are consistent with an expected scale height of about 50 Mm. For the second case a reasonably confined estimate for the density scale height in between 20 and 99 Mm is calculated. For a different event and following the same method, \cite{vandoorsselaere07} obtain a value of 109 Mm, which is about double the estimated hydrostatic value.

The period ratio approach has also been followed by \cite{diaz10}  to obtain information about the density structuring along prominence threads. These authors showed that the dimensionless oscillatory frequencies of the fundamental kink mode and the first overtone are almost independent of the ratio of the thread diameter to its length. Thus, the dependence on the length of the tube and the thread Alfv\'en speed can be removed by considering the period ratio,
\begin{eqnarray}
\label{diazcurves}
\frac{P_1}{2 P_2} = F (W/L, \rhop/\rhoc).
\end{eqnarray}
Equation~(\ref{diazcurves}) can be used for diagnostic purposes, once reliable measurements of multiple mode periods are obtained.  From an observational point of view, there seem to be hints of the presence of multiple mode oscillations in observations by  \cite{lin07}, who reported on the presence of two periods, $P_1=16$~min and $P_2=3.6$~min in their observations of a prominence region. Reference \cite{diaz10} used the period ratio from these observations to infer a value of  $\vap\sim160$~km s$^{-1}$ for the prominence Alfv\'en speed.

%%%%%%%%%%%%%%%%%%%%%%%%%%%%%
\section{Seismology Using the Damping of Oscillations}
%%%%%%%%%%%%%%%%%%%%%%%%%%%%%

Soon after transverse coronal loop oscillations were discovered a number of physical mechanisms were proposed to explain their quick time damping. We concentrate here on resonant absorption (\cite{ruderman02,goossens02}).  Damping is another source of information for plasma diagnostics and seismology using resonant absorption enables us to infer information about the transverse density structuring in magnetic flux tubes. The period and damping ratio of kink oscillations in the thin tube and thin boundary limits can be expressed by the following analytic approximations

\begin{equation}
P  =  \tau_{\rm Ai} \; \sqrt{2} \; \; \left(\frac{\zeta + 1}{\zeta}\right)^{1/2} \label{T} \mbox{\hspace{.5cm}} \mbox{and} \mbox{\hspace{.5cm}}  \frac{\displaystyle \tau_{\mathrm d}}{\displaystyle P} = \frac
{\displaystyle 2} {\displaystyle \pi} \frac{\displaystyle \zeta +
1}{\displaystyle \zeta - 1} \frac{\displaystyle 1}{\displaystyle
l/R}.\label{forward}
\end{equation}
These equations express the period, $P$ and the damping time, $\tau_{\rm d}$, which are observable quantities 
in terms of the internal Alfv\'en travel time, $\tau_{\rm Ai}$, the density contrast, $\zeta=\varrho_{\rm i}/\varrho_{\rm e}$, and the transverse inhomogeneity length scale, $l/R$, in units of the radius of the loop. By only considering
the damping ratio,  \cite{ruderman02} estimated a transverse inhomogeneity of $l/R=0.23$, for an event observed by \cite{nakariakov99} after assuming a density contrast of 10. The determination of transverse density structuring in a set of observed coronal loop oscillations was performed by \cite{goossens02} and values of the transverse inhomogeneity length-scale in between 0.15 and 0.5 were obtained.  Eigenmode computations and comparison to observations for highly inhomogeneous loops were performed by \cite{vandoorsselaere04,aschwanden03b}.

%This result indicated that the thin boundary approximation ($l \ll R$) used to derive the analytical damping ratio might be outside the range of validity for a real application. Reference \cite{vandoorsselaere04}, numerically computed damping rates  outside the thin tube and thin boundary approximation  that were subsequently used for an observational test of resonant absorption by \cite{aschwanden03b}. 

The first seismology inversions that used the observational information on both periods and damping times in the context of resonant damping in a consistent manner were performed by \cite{arregui07,goossens08a}. Their important finding is that when no further assumptions are made on any of the unknown parameters, the inversion gives rise to a one-dimensional solution curve in the three-dimensional parameter space (see Fig.~\ref{fig:seismology}a). Although there is an infinite number of solutions that equally well reproduce observations, the internal Alfv\'en travel time is constrained to a narrow range. Further information on coronal loop seismology using resonantly damped oscillations can be found in the review by \cite{goossens08b}.

Prominence thread oscillations also display damping (\cite{lin09}) and damping by resonant absorption provides a plausible explanation (\cite{arregui08}). The period and damping time of thread oscillations are seen to depend on the length of the thread ($L_p/L$) in units of the length of the magnetic tube (\cite{soler10,arregui11b}), but the damping ratio $P/ \tau_{\rm d}$ is almost insensitive to this parameter.
When applying the period and damping inversion technique to prominence threads, one inversion curve is obtained for each value of the length of the thread.  The solutions obtained by \cite{soler10}  for a set of values for $L_p$ are shown in Fig.~{\ref{fig:seismology}b.  Because of the insensitiveness of the damping ratio with density contrast ($\rho_{\rm p}/\rho_{\rm c}$), in prominence plasmas,  the obtained solution curves display an asymptotic behavior for large values of density contrast that enables us  to obtain precise estimates for the thread Alfv\'en speed and the transverse inhomogeneity length scale. 

Seismology inversions using the damping of oscillations have also been performed by interpreting the damping of vertically polarized  kink oscillations in coronal loops by wave leakage. Fast waves with frequencies above the external Alfv\'en frequency cannot be trapped and radiate energy away the structure. Reference \cite{verwichte06}
obtain, from the observed oscillation period, damping time and relative intensity amplitude, the values for the loop density contrast, the local slope of the Alfv\'en frequency, and the Alfv\'en speed at the loop axis self-consistently. A shortcoming of the curved slab model is that it predicts a wave leakage that is too strong to explain the observations.

\begin{figure*}[t]
\centering
  \includegraphics[height=4.4cm,width=5.0cm]{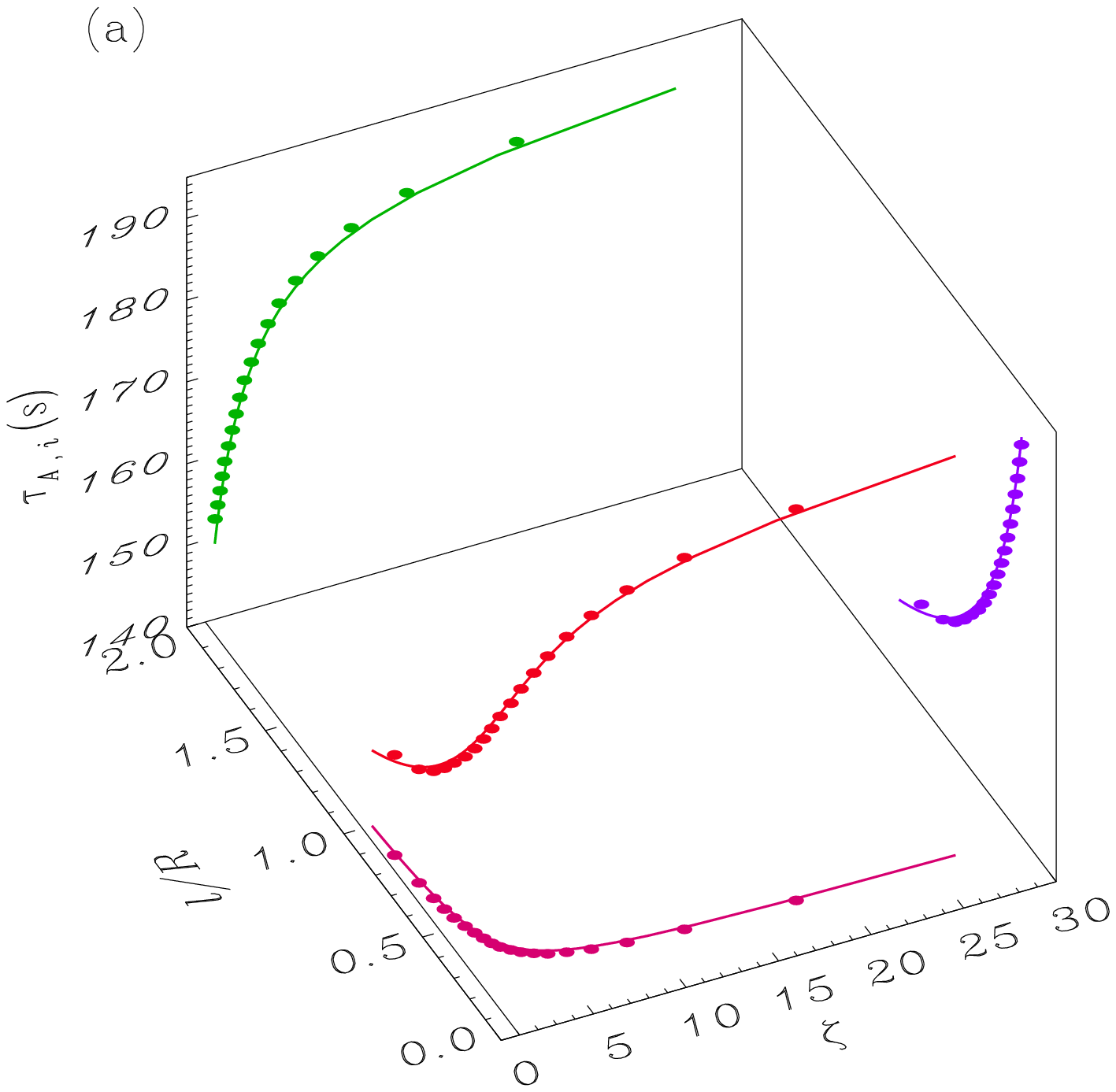} 
   \includegraphics[height=4.4cm,width=5.0cm]{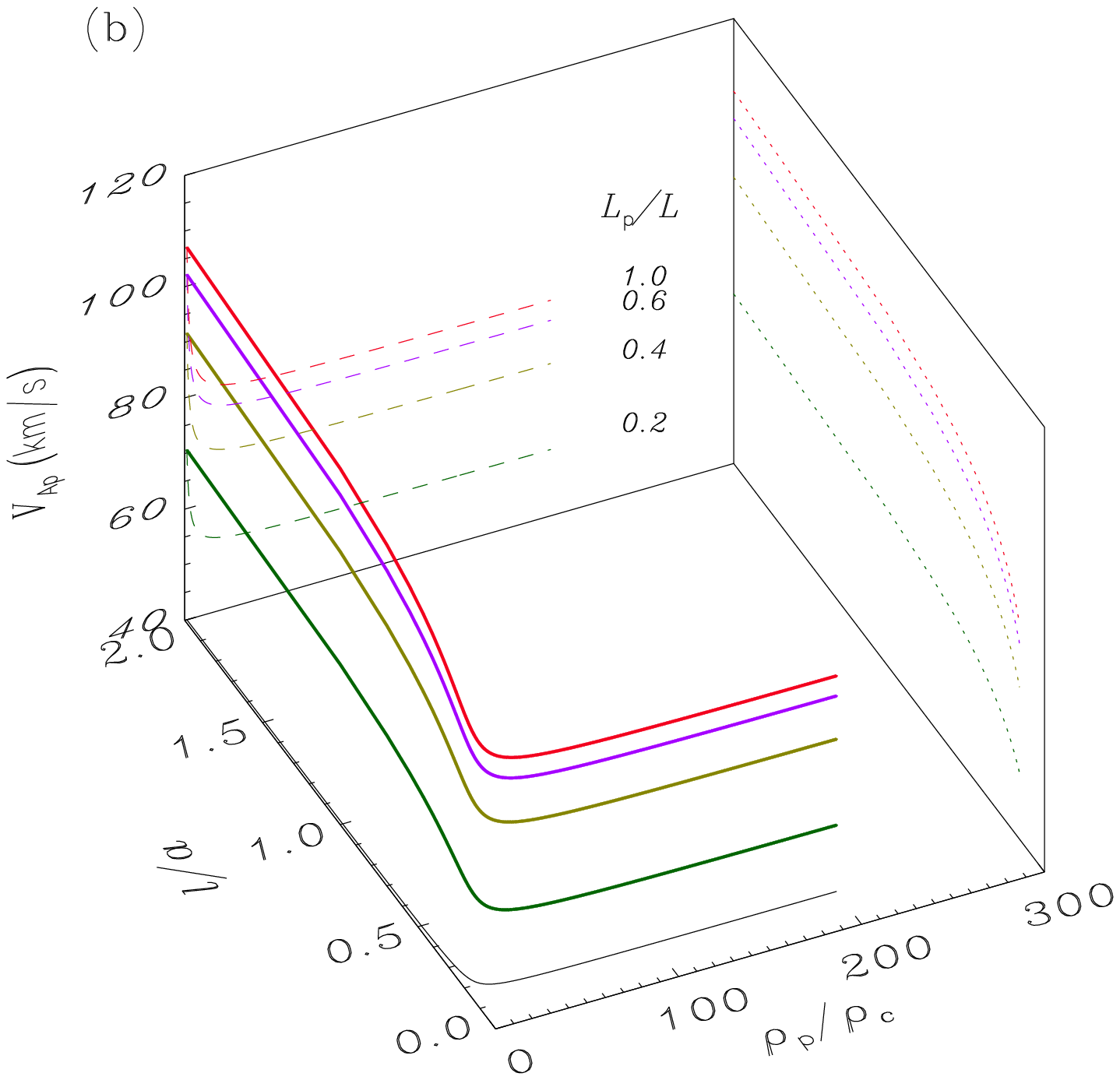}	
   \caption{{\bf (a)} Inversion curve for a coronal loop oscillation with $P=272$ s and $\tau_{\rm d}=849$ s in the parameter space of unknowns. {\bf (b)} Application of the same inversion technique to a prominence thread using different lengths. The observed period and damping time are  20 and 60 minutes, respectively, and $L=10^5$ km.}
   \label{fig:seismology}
\end{figure*}

%%%%%%%%%%%%%%%%%%%%%%%%%%%%%%%%%%%%%%%%
\section{Seismology in the Presence of Flows}
%%%%%%%%%%%%%%%%%%%%%%%%%%%%%%%%%%%%%%%%

%Material flows are observed everywhere in the solar atmosphere. Two recent studies have incorporated flows in the inversion of physical parameters in coronal loops and active region prominence threads, respectively.

The first  seismological application of prominence seismology using Hinode observations of flowing and transversely oscillating threads was presented by \cite{terradas08}, using observations obtained in an active region filament by \cite{okamoto07}. The observations show a number of  threads that flow following a path parallel to the photosphere while they are oscillating in the vertical direction. Reference \cite{terradas08} interpret these oscillations in terms of the kink mode of a magnetic flux tube. First, by neglecting the effect of flows, \cite{terradas08} find that a one-to-one relation between the thread Alfv\'en speed and the coronal Alfv\'en speed can be established. For one of the observed threads, and  considering a length of the total magnetic flux tube of $L=100$ Mm, an overall value for the thread Alfv\'en speed between 120 and 350 km s$^{-1}$ is obtained. Next, mass flows are considered and \cite{terradas08}  find that the flow velocities measured by \cite{okamoto07} result in slightly shorter (3\% to 5\%) kink mode periods than the ones derived in the absence of flow. Hence, on this particular case, the detected flow speeds would produce only slightly different results in the seismology inversion.

More recently, \cite{terradas11} argue that the presence of siphon flows can cause an underestimation of magnetic field strength in coronal loops using the traditional  method outlined in Sect.~\ref{period}. In particular, the calculation of the kink speed and the estimation of magnetic field strength, assuming a static model,  give values that are considerably smaller than the ones obtained in the presence of flow. 
The reason is that, contrary to the static case, different positions along the tube oscillate with a different phase. The theory is applied to the linear phase shift reported along the post-flare coronal loop analyzed by \cite{verwichte10}, by assuming that its cause is  a siphon flow. The inversion result shows that the flow would be in the fast flow regime ($\sim$10$^{3}$ km s$^{-1}$). This is not completely unreasonable in the dynamic post-flare environment of coronal loops.

\section{Seismology in the Spatial Domain}\label{spatial}

For longitudinally stratified loops, \cite{andries05a} noticed that the amplitude profiles for the perturbed variables along the magnetic field direction are directly affected by  density stratification.  Several studies make use of observations of the spatial distribution of equilibrium model parameters and  wave properties for plasma diagnostic purposes. The first studies that make use of spatial information from observations, in combination with theoretical results of the effect of density stratification are by \cite{erdelyi07,verth07b}. These authors show that density stratification causes the anti-nodes of the first harmonic of the standing kink mode in coronal loops to shift towards the loop foot-points.  The anti-node shift corresponding to a density scale height of 50 Mm and a loop half length of 100 Mm is approximately 5.6 Mm. Shifts in the Mm range are measurable quantities, hence spatial seismology  was proposed as a complementary method of probing the plasma stratification in the corona. 

Magnetic field inhomogeneity also affects oscillation properties.  A varying longitudinal magnetic field implies an equilibrium with flux tube expansion. This expansion is a measurable quantity. Reference  \cite{verth08a} combine density and magnetic field structuring and,  in the context of seismology using period ratios,  find that even a relatively small coronal loop expansion can have a significant  effect on the inferred density scale height. A detailed and step-by-step inversion is presented by  \cite{verth08b}. The results indicate that using the observed ratio of the first overtone and fundamental mode to predict the plasma density scale height and not taking account of loop expansion will lead to an overestimation of scale heigh by a factor of 2. Similar physical models and inversion techniques have recently been applied to other wave types, e.g., Alfv\'en waves in stratified waveguides (\cite{verth10}), and to different magnetic and plasma structures, such as spicules (\cite{verth11}).

\begin{figure}[t]
\sidecaption
  \includegraphics[height=4.0cm,width=5.0cm]{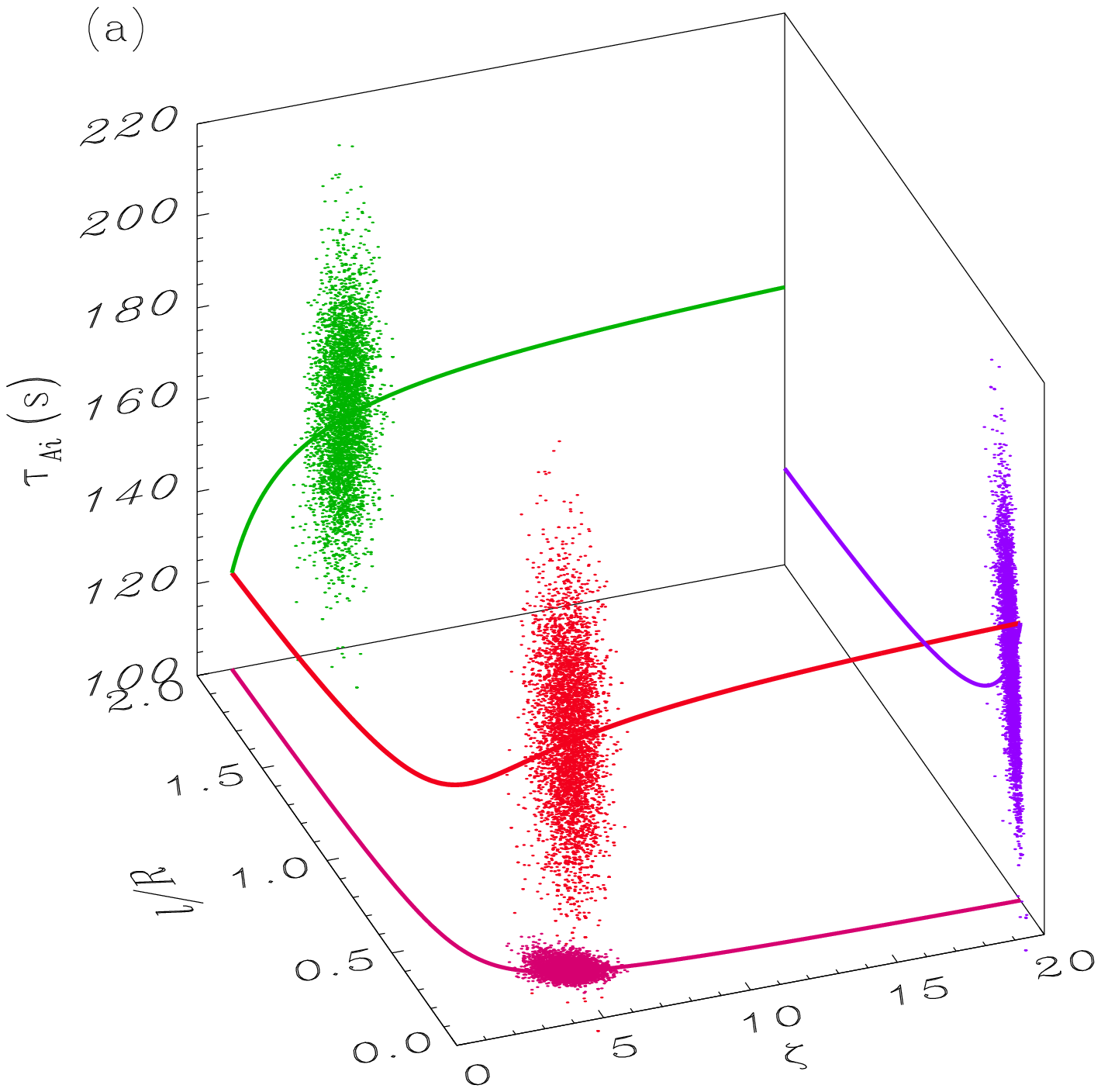} 
   \includegraphics[height=4.0cm,width=5.0 cm]{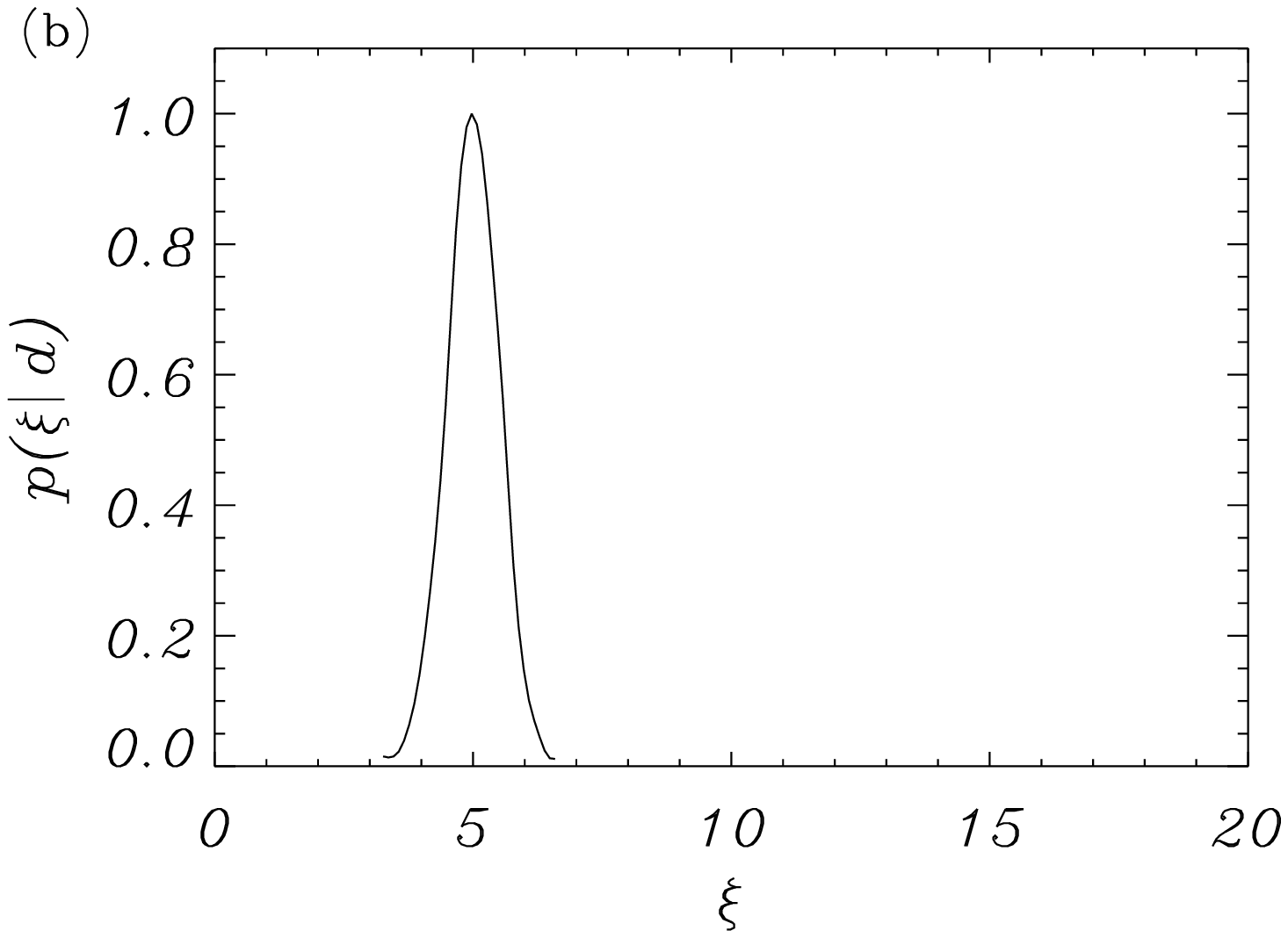}\\
   \includegraphics[height=4.0cm,width=5.0cm]{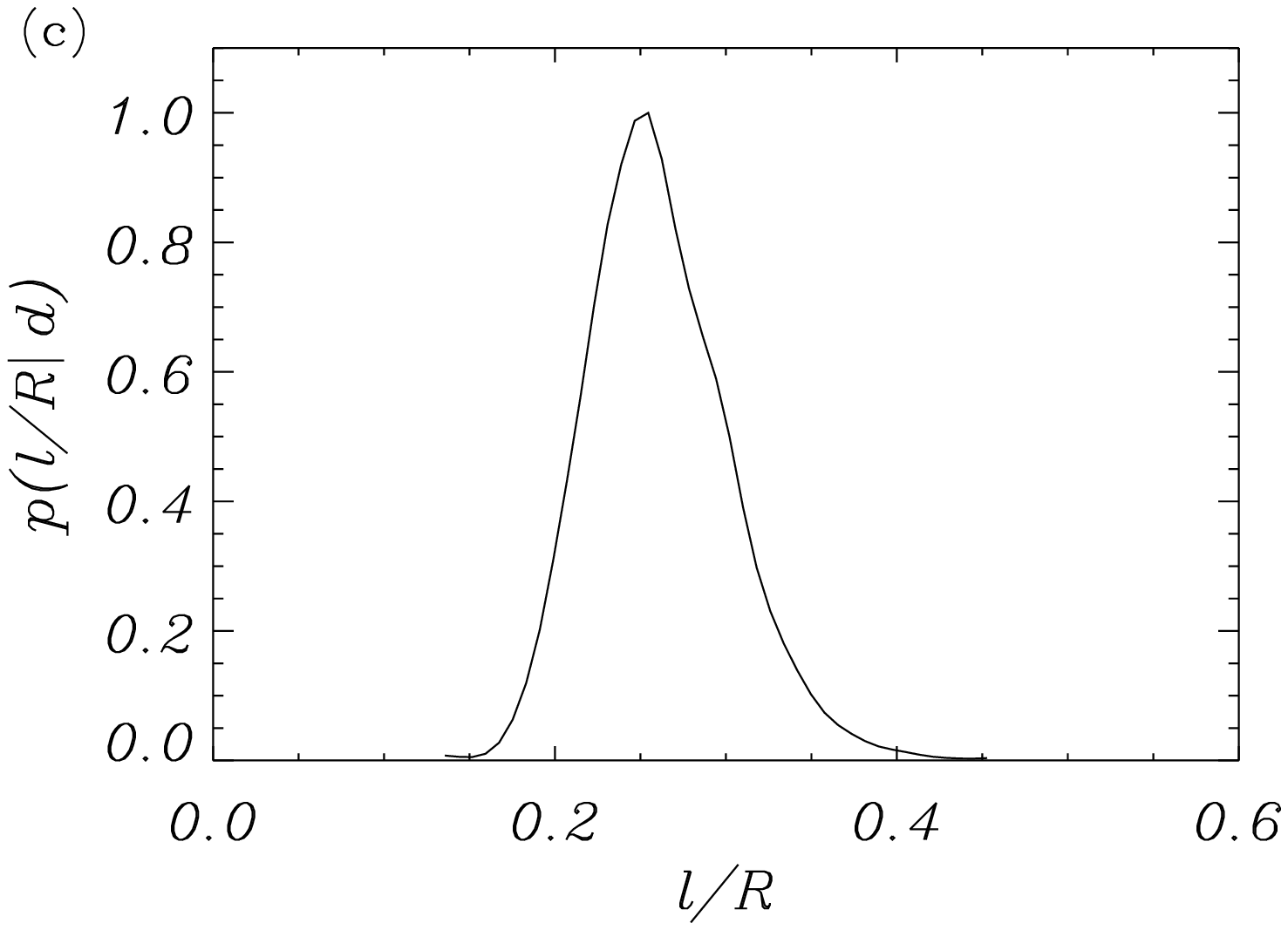} 
   \includegraphics[height=4.0cm,width=5.0cm]{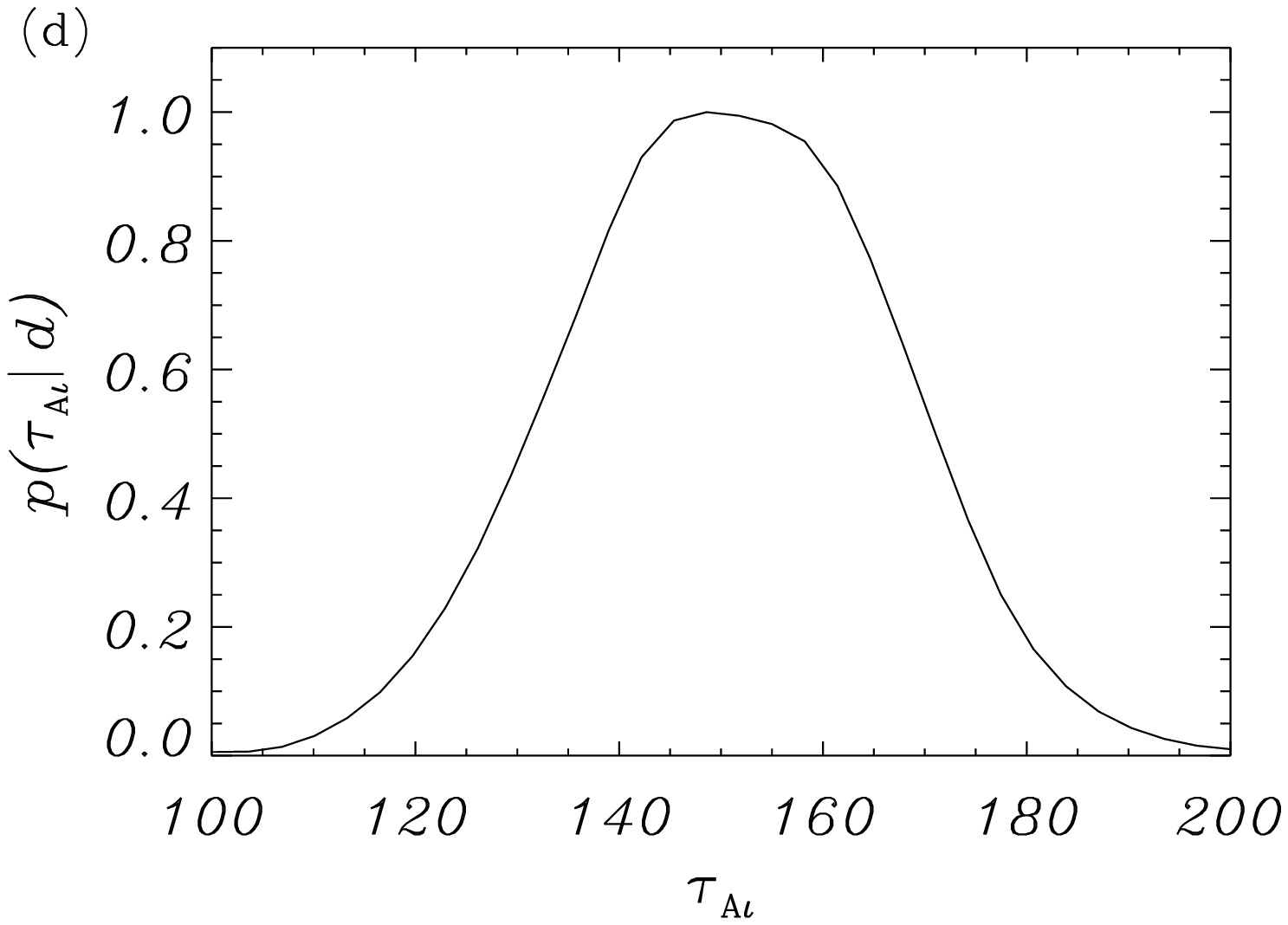}
   \caption{{\bf (a)} Bayesian inversion technique results in the form of converged Markov chain elements in the parameter space, together with analytic inversion result in solid line.  {\bf (b)--(c)} Marginal posteriors for density contrast, transverse inhomogeneity and Alfv\'en travel time. The observed period and damping time are  232 s and  881 s, respectively.}
   \label{fig:bayesian}
\end{figure}

\section{Bayesian Inference}

Reference \cite{arregui11b}  have recently proposed an alternative statistical approach to the inversion problem, based on parameter inference in the bayesian framework. The method makes use of the Bayes' theorem. According to this rule, the state of knowledge on a given set of parameters (the posterior distribution), is a function of what is known a priori, independently of the data, (the prior), and the likelihood of obtaining a data realization actually observed as a function of the parameter vector (the likelihood function). 
%Bayes' theorem defines the inversion problem and the computation of the posterior is the solution to the problem, since it is inference conditional on observed data. The most plausible model is the one that maximizes the posterior. 
In an application to transverse coronal loop oscillations, posterior probability distribution functions were obtained by means of Markov Chain Monte Carlo simulations, incorporating observed uncertainties in a consistent manner. By using observational estimates for the density contrast by \cite{aschwanden03b},  \cite{arregui11b} find well-localized solutions in the posterior probability distribution functions for the three parameters of interest (see Fig.~\ref{fig:bayesian}).  From these probability distribution functions, numerical estimates for the unknown parameters can be obtained. The uncertainties on the inferred parameters are given by error bars correctly propagated from observed uncertainties.

\section{Conclusion}

The application of seismology techniques to better understand the physics of  the solar atmosphere has produced fruitful results in the last years. Some relevant examples are mentioned in this paper.
Nearly all the necessary steps for a proper seismology seem to have been  followed, i.e., 
the observational evidence of oscillations and waves;  the correct identification of observed with theoretical wave modes;  the determination or restriction of physical parameters and their structuring.  
The challenge that is recurrently mentioned is the need for better observations and more realistic analytic/numerical models. Besides this, often the quantity of unknowns outnumbers that of measured wave properties and information is always incomplete and uncertain. In this respect, the use Bayesian parameter inference methods could be of high value for the future of solar atmospheric seismology, when the inversion of physical parameters will be based on the combination of large scale numerical parametric results and the analysis of data sets obtained from, e.g., SOLAR-C.

\begin{acknowledgement}
The author acknowledges  the financial support received from the Spanish MICINN and FEDER funds under Grant No.\ AYA2006-07637. The author is also grateful to the Solar Physics Group members at Universitat de les Illes Balears, for many years of fruitful work. 
\end{acknowledgement}
%

%\bibliographystyle{spmpsci}
%\bibliography{seismology}

%\input{referenc}
\end{document}